\documentclass[aps,prd,twocolumn,a4paper]{revtex4-1}
\usepackage{ulem}
\usepackage{color}
\usepackage{graphicx}
\usepackage{hyperref}   
\usepackage{appendix}
\usepackage{epsfig}
\usepackage{epstopdf}
\usepackage{amsmath}
\usepackage{subfig}
\usepackage{comment}
\usepackage[dvipsnames]{xcolor}
\newcommand{\be}{\begin{equation}}
\newcommand{\ee}{\end{equation}}
\newcommand{\ba}{\begin{eqnarray}}
\newcommand{\ea}{\end{eqnarray}}

\begin{document}
\title{Longitudinal conductivity of hot magnetized collisional QCD medium in the inhomogeneous electric field}
\author{Manu Kurian }
\email{manu.kurian@iitgn.ac.in}
\author{Vinod Chandra }
\email{vchandra@iitgn.ac.in}
\affiliation{Indian Institute of Technology Gandhinagar,  Gandhinagar-382355, Gujarat, India}

\begin{abstract}
The longitudinal current density induced by the inhomogeneous electric field in the hot magnetized 
quark-gluon plasma has been investigated and utilized in obtaining the conductivity 
of the medium. The analysis has been done in the regime 
where inhomogeneity of the field is small so that the collision effect could be significant. 
The modeling of the QCD medium is based on a quasiparticle description where the medium 
effects have been encoded in the effective quarks, antiquarks and gluons. The temperature 
dependence of the linear longitudinal current density (in terms of the electric field) and 
the additional components of current density due to the inhomogeneity of electric field 
(in terms of its derivatives) have been obtained by solving the $(1+1)-$dimensional effective 
covariant kinetic theory with a proper collision term. The conductivity has been obtained from the 
current density in the presence of the inhomogeneous field. The collisional aspects of the medium have 
been captured by including both thermal relaxation approximation and the Bhatnagar-Gross-Krook collision 
kernels in the analysis. Further, the hot QCD medium effects and higher 
Landau level contributions to the current density and the conductivity have been investigated. 
It has been seen that the effects of inhomogeneity of the field and the mean field corrections to the current density 
and the conductivity are more visible in the temperature regions which are not far from the transition temperature.  
\\ 

\end{abstract}

\maketitle
 
\section{Introduction}
Quantum chromodynamics (QCD) predicts a transition from normal hadronic matter 
to a deconfined state of matter called quark-gluon-plasma (QGP). 
The relativistically energetic heavy-ion collision 
experiments at Relativistic heavy-ion collider (RHIC) and large hadron collider (LHC), 
verified the existence of the QGP, as a near-perfect fluid~\cite{STAR, Aamodt:2010pb}. Intense magnetic field has been
 created in the early stages of non-central asymmetric heavy-ion collisions and has great phenomenological 
significance~\cite{Skokov:2009qp,deng,Zhong:2014cda,Das:2016cwd, Ghosh:2018xhh,Mukherjee:2017dls, 
Karmakar:2018aig,Das:2017vfh,Karmakar:2019tdp,
Bandyopadhyay:2017cle,Koothottil:2018akg,Roy:2017yvg,Coci:2019nyr}. Chiral magnetic 
effect~\cite{Fukushima:2008xe,Sadofyev:2010pr,Huang,She:2017icp}, 
anomalous charge separation~\cite{Huang:2015fqj}, chiral vortical 
effect~\cite{Kharzeev:2015znc,Avkhadiev:2017fxj,Yamamoto:2017uul} 
and more recently, the realization  of global $\Lambda$-hyperon polarization 
in RHIC~\cite{STAR:2017ckg,Becattini:2016gvu} are some of the interesting aspects 
that are focus areas of current research on strong interaction physics at the RHIC. In addition, 
the intense magnetic field may affect 
the transport and thermodynamic properties of the QGP~\cite{Hattori:2017qih,Hattori:2016lqx,Kurian:2018qwb,
Kurian:2017yxj,Hattori:2016cnt,Fukushima:2017lvb}. 
  
The quark and antiquark dynamics is described by $(1+1)-$dimensional Landau level (LL)
kinematics (dimensional reduction) in the presence of the magnetic field~\cite{Hattori:2017qih,
Kurian:2018dbn}. The gluonic dynamics in the magnetized QGP can be 
indirectly affected through self-energy where quark/antiquark loop contributions are 
taken into account. The microscopic properties of Landau level 
 dynamics of quarks, antiquarks result induce significant modifications 
to the macroscopic transport properties~\cite{Hattori:2017qih,
Kurian:2018dbn}. In the Refs~\cite{Rath:2019vvi,Li:2018ufq,Feng:2017tsh,
Mohanty:2018eja,Huang:2011dc,Critelli:2014kra}, the authors have 
investigated various transport coefficients in the presence of the magnetic field. Among them,
conductivity is the most significant one, that describes the electromagnetic responses 
of the medium and controls the late time behaviour of electromagnetic fields. 
Inhomogeneous electromagnetic fields are expected to be generated in the 
heavy-ion collider experiments~\cite{deng,Tuchin:2013apa,Hongo:2013cqa,Tuchin:2012mf}. 
The simplest component of current which is linear to the electric field is the Ohmic current. 
Since the inhomogeneity of the electromagnetic field has been already realized in 
heavy-ion collision experiments, it is an interesting task to investigate the additional
components of the current due to the spacetime inhomogeneities of fields 
that are not negligible compared with the linear component while 
including the collisional effects. The longitudinal conductivity could be 
defined from the current itself. This sets the motivation for the current analysis.

The fluctuations of the electromagnetic field in the
heavy-ion collision experiments might play significant
role in understanding the electromagnetic responses of the medium. 
Recent investigations~\cite{Zakharov:2017yst,Tuchin:2013ie} reveal that the
fluctuations are much smaller as compared to the earlier
predictions. In the current analysis, we are focusing on the regime where 
inhomogeneity of the electromagnetic field is small so that the collision effects cannot be 
neglected~\cite{Satow:2014lia}. For this purpose, proper collision integral for the processes 
that are dominant in the presence of a strong magnetic 
field need to be incorporated in the effective kinetic theory. The collisional aspects can be 
included either through relaxation-time approximation (RTA) or  Bhatnagar-Gross-Krook (BGK) kernel. 
Recall that the quark-antiquark pair production and fusion processes are kinematically 
possible in the strong magnetic field limit~\cite{Hattori:2016lqx}. These processes are dominant 
over $2\rightarrow 2$ processes because of the fact that rate of the former one is 
proportional to the QCD running coupling constant $\alpha_{s}$ whereas that of binary processes is 
proportional to $\alpha_s^2$. In Ref.~\cite{Kurian:2018qwb}, 
authors computed the momentum dependent 
thermal relaxation time for the $1\rightarrow 2$ processes in the presence of the magnetic 
field. The collisional effects can also be described through BGK kernel. The difference 
between the BGK collisional term and the conventional relaxation RTA is that, the former case the number of 
particles is instantaneously conserved. Refs.~\cite{Hattori:2016lqx,Kurian:2017yxj,Hattori:2016cnt}
describes the estimation of longitudinal current density (longitudinal conductivity) in the 
leading order in the presence of the strong magnetic field within the regime $T^{2}\ll \mid q_fB\mid$.
Since the magnetic field is considerably strong in this regime, the thermal occupation 
of the higher Landau levels (HLLs) is suppressed exponentially as $e^{-\frac{\sqrt{q_fB}}{T}}.$
This allows to focus only on the lowest 
Landau level (LLL) state of quarks and antiquarks in the regime $T^{2}\ll \mid q_fB\mid$.
Recent investigations~\cite{Fukushima:2017lvb, Kurian:2018qwb} showed that the impact of 
HLLs are significant for the transport phenomena for an arbitrary magnetic field. The current analysis is on a 
more realistic regime $gT\ll \sqrt{\mid q_fB\mid}$ with full Landau level resummation.

 The prime focus of the present analysis is to estimate the longitudinal current density of 
 the magnetized QGP including the effects of inhomogeneity of electric field and HLLs, with proper collision term while 
incorporating the hot QGP medium interactions/effects. The analysis is done with the 
relativistic covariant kinetic theory by employing the effective fugacity quasiparticle 
model (EQPM)~\cite{Chandra:2011en,Mitra:2018akk}. Hot QCD medium effects are incorporated in 
the EQPM quark/antiquark and 
gluonic degrees of freedom considering QGP as a Grand-canonical ensemble. 
 Here, the current density picks up the 
mean field contribution from the local particle flow and 
energy-momentum conservation conditions within the effective kinetic 
theory. The mean field correction to the transport coefficients such as shear viscosity, 
bulk viscosity and thermal conductivity of a hot magnetized QGP is estimated in our 
previous work~\cite{Kurian:2018qwb}. In addition, one could read off the longitudinal conductivity from 
the current density. The temperature behaviour and magnetic field dependence of conductivity have 
also been investigated.

The manuscript is organized as follows. In section II, the mathematical formalism 
of the estimation of the longitudinal current density and conductivity in the hot QGP medium with the 
RTA and BGK kernel within the effective covariant kinetic 
theory is presented. Predictions of the effects due to inhomogeneity of field and HLLs 
in the current density and conductivity 
along with the mean 
field contributions are discussed in section III. Finally, in section IV the conclusion 
and outlook of the article are presented.     
 
 \section{Longitudinal conductivity of the QGP in the magnetic field}
 
The inhomogeneity of the electric field generates the components 
of current in terms of spacetime derivatives of the field.
The formalism for the estimation of the current density 
include the quasiparticle modeling of the QGP 
away from the equilibrium followed by the 
setting up of the effective covariant kinetic theory for 
the processes under consideration. The medium interactions are encoded 
in the quasiparticle models either through effective fugacity or effective mass parameter.
The current analysis is done within the effective fugacity 
quasiparticle model (EQPM)~\cite{Chandra:2011en, Mitra:2018akk, Chandra:2007ca,Mitra:2017sjo}, that encodes the 
($2+1)-$flavor lattice equation of state (EoS)~\cite{Cheng:2007jq,Borsanyi}. 
The temperature dependent quark and gluonic effective 
fugacities, $z_{q/\bar{q}}$ and $z_g$ respectively, describe the hot QCD medium interaction effects in 
the system. The extension of EQPM in the presence of the strong magnetic field is 
described in the Ref.~\cite{Kurian:2017yxj}.

For setting up the covariant effective kinetic theory within EQPM~\cite{Mitra:2018akk} in the 
 presence of the strong magnetic field, we first need to define the basic macroscopic 
 quantities of the system. We begin with the energy-momentum tensor $T^{\mu\nu}$ 
 and particle four-flow  $N^{\mu}(x)$ in the magnetized medium.
The equilibrium energy-momentum tensor (longitudinal) in the 
presence of the strong magnetic field $\vec{B}=B\hat{z}$ can be defined in terms 
of quasiparticle momenta as follows~\cite{Kurian:2018qwb},
\begin{align}\label{5}
T^{\mu\nu}(x)&=\sum_{l=0}^{\infty}\sum_{k=q,\bar{q}}\dfrac{\mid {q_f}_kB\mid}{2\pi}\mu_l
N_c\int_{-\infty}^{\infty}{\dfrac{d{p}_{z_k}}{(2\pi)\omega_{l_k}}
{p}_k^{\mu}{p}_k^{\nu}f^0_k(x, {p}_{z_k})}\nonumber\\
&+\sum_{l=0}^{\infty}\sum_{k=q,\bar{q}}\delta\omega\dfrac{\mid {q_f}_kB\mid}{2\pi}\mu_l
N_c\int_{-\infty}^{\infty}{\dfrac{d{p}_{z_k}}{(2\pi)
{\omega_{l_k}}}\dfrac{\langle{p}_k^{\mu}{p}_k^{\nu}\rangle}
{E_{l_k}} }\nonumber\\
&\times f^0_k(x,{p}_{z_k}).
\end{align}
In the local rest frame in which the hydrodynamic four-velocity  
$u^{\mu}=(1, {\bf 0})$,  the quasiquark/antiquark momentum distribution function in the strong magnetic field has the 
following form,
\begin{equation}\label{2}
f^0_{q}=\dfrac{z_{q}\exp{[-\beta (u^{\mu}\bar{p}_{\mu})]}}{1+ z_{q}\exp{[-\beta (u^{\mu}\bar{p}_{\mu})]}}.
\end{equation}
The quasiparticle dressed momenta $p^{\mu}$ can be defined in terms of bare particle momenta $\bar{p}^{\mu}$ 
through the dispersion relations as,
\begin{align}\label{3}
p^{\mu}&=\bar{p}^{\mu}+\delta\omega u^{\mu}\equiv(\omega_{l},0,0,p_z),
& \delta\omega= T^{2}\partial_{T} \ln(z_{q}).
\end{align}
The dispersion relation that
encodes the collective excitation of 
quasiparton modifies the zeroth component of the 
quasiparticle four-momenta in the local rest frame as,
\begin{equation}\label{7.0}
\omega_{l}=E_{l}+\delta\omega,
\end{equation} 
in which $E_{l}\equiv\sqrt{p_{z}^{2}+m^{2}+2l\mid q_fB\mid}$ is the 
Landau levels of the quarks/antiquarks in the presence of magnetic field
($m_u = 3$ MeV,
$m_d = 5$ MeV and $m_s = 100$ MeV are the up, down and strange quarks masses respectively). 
Since the fermionic dynamics gets constrained to $(1+1)-$dimensional space,
the longitudinal projection operator $\Delta_{\|}^{\mu\nu}$ in the dimensionally reduced space takes 
the form~\cite{Li:2017tgi},
\begin{equation}\label{4}
\Delta_{\|}^{\mu\nu}\equiv g_{\|}^{\mu\nu}-u^{\mu}u^{\nu},
\end{equation}
with $g_{\|}^{\mu\nu}=(1,0,0,-1)$. 
Here, $\langle p_k^{\mu} p_k^{\nu}\rangle=
\dfrac{1}{2}(\Delta_{\|}^{\mu\alpha}\Delta_{\|}^{\nu\beta}+
\Delta_{\|}^{\mu\beta}\Delta_{\|}^{\nu\alpha})p_{\alpha}p_{\beta}$.  
The integration phase factor in the strong magnetic field 
due to the dimensional reduction~\cite{Bruckmann:2017pft,Tawfik:2015apa,
Gusynin:1995nb} is defined as, 
\begin{equation}\label{7}
\int{\dfrac{d^{3}p}{(2\pi)^{3}}}\rightarrow
\dfrac{\mid q_fB\mid}{2\pi}\int_{-\infty}^{\infty}{\dfrac{dp_{z}}{2\pi}}\mu_l,
\end{equation}   
with the Landau level degeneracy factor $\mu_l=(2-\delta_{l0})$.
The energy-momentum tensor Eq.~(\ref{5}) gives the exact form of longitudinal pressure 
($P_{\|}$) and energy density ($\varepsilon_{\|}$)
in the strong magnetic field as described in~\cite{Kurian:2017yxj}, from the macroscopic description,
\begin{equation}\label{8}
T^{\mu\nu}=\varepsilon_{\|} u^{\mu}u^{\nu}-P_{\|}\Delta_{\|}^{\mu\nu},
\end{equation}
from which $\varepsilon_{\|}=u_{\mu}u_{\nu}T^{\mu\nu}$ and $P_{\|}= 
{\Delta_{\|}}_{\mu\nu}T^{\mu\nu}$ can be defined.

Following the same arguments the particle four flow $N^{\mu}$ of the magnetized medium 
takes the following form,
\begin{align}\label{9}
N^{\mu}(x)&=\sum_{l=0}^{\infty}\sum_{k=q,\bar{q}}\dfrac{\mid {q_f}_kB\mid}{2\pi}\mu_l
N_c\int_{-\infty}^{\infty}{\dfrac{d{p}_{z_k}}{(2\pi)\omega_{l_k}}
{p}_k^{\mu}f^0_k(x,{p}_{z_k})}\nonumber\\
&+\sum_{l=0}^{\infty}\sum_{k=q,\bar{q}}\delta\omega\dfrac{\mid {q_f}_kB\mid}
{2\pi}\mu_lN_c\int_{-\infty}^{\infty}{\dfrac{d{p}_{z_k}
}{(2\pi){\omega_{l_k}}}\dfrac{\langle{p}_k^{\mu}\rangle}
{E_{l_k}}}\nonumber\\
&\times f^0_k(x,{p}_k),
\end{align}
with $\langle p^{\mu}\rangle={\Delta_{\|}}^{\mu\nu}
p_{\nu}$. The zeroth component of the $N^{\mu}$  gives the expression of 
number density $n$ for quarks and antiquarks as,
\begin{equation}\label{10}
n=\sum_{l=0}^{\infty}\sum_{k=q,\bar{q}}\dfrac{\mid {q_f}_kB\mid}{2\pi} \mu_lN_c \int_{-\infty}^{\infty}{\dfrac{dp_{z}}{(2\pi)}}f_{k}^{0},
\end{equation}  
which reduced to the LLL result as described in the Ref.\cite{Kurian:2017yxj},
$i.e.,$ $n=\sum_{k=q,\bar{q}}\dfrac{\mid {q_f}_kB\mid}{2\pi}N_c\ln(1+z_{q})$ 
within the regime $T^{2}\ll \mid q_fB\mid$.
Here, we need to calculate the induced longitudinal current density of the magnetized QGP in the 
presence of an external electric field $\vec{E}=E(X)\hat{z}$. 
 In the current analysis, we are focusing on the HLLs contribution of quarks and antiquarks 
to the linear and nonlinear components of the longitudinal current density in the presence 
of the magnetic field $\vec{B}=B\hat{z}$. The authors of the Ref.~\cite{Fukushima:2017lvb} 
showed that hall conductivity (transverse component) vanishes in the one-loop order of 
polarization tensor from the Landau quantization of transverse motion.
Hence, we have,
\begin{align}\label{11}
\vec{J}&=\sum_{l=0}^{\infty}q_f\dfrac{\mid q_fB\mid}{2\pi}\mu_lN_c\int_{-\infty}^{\infty}{\dfrac{dp_z}{2\pi} \vec{v}(f_q-f_{\bar{q}})}\nonumber\\
&-\delta\omega\sum_{l=0}^{\infty}q_f\dfrac{\mid q_fB\mid}{2\pi}\mu_lN_c\int_{-\infty}^{\infty}{\dfrac{dp_z}{2\pi} 
\dfrac{\vec{v}}{E_{l}}(f_q-f_{\bar{q}})},
\end{align}
where $v_z=\frac{p_z}{\omega_{l}}$ is 
the longitudinal velocity. The second term of the Eq.~(\ref{11}) describes the mean field contribution 
to the current density within EQPM. The local momentum distribution function of quarks can be expand as 
\begin{equation}\label{12}
f_{q/\bar{q}}= f^{(0)}_{q/\bar{q}}(p_z)+\delta f^{(1)}_{q/\bar{q}}(p_z,X)+O(F^2_{\mu\nu}),
\end{equation}
where  $X=(X_0=t, \vec{X}=z\hat{z})$ and the equilibrium EQPM quasi-quark/antiquark distribution function is given as,
\begin{equation}\label{13}
f^0_{q/\bar{q}}=\dfrac{z_{q}\exp{(-\beta \sqrt{p_{z}^{2}+m^{2}+2l\mid q_fB\mid
})}}{1+ z_{q}\exp{( -\beta 
\sqrt{p_{z}^{2}+m^{2}+2l\mid q_fB\mid} )}}.
\end{equation}
The quantity $\delta f^{(1)}_q$ is the change from 
the local distribution function. Hence, the determination of the longitudinal current 
density in the strong magnetic field requires the knowledge of the system away 
from the equilibrium. The dynamics of the distribution function in the 
strong field limit can be explicitly described by the $(1+1)-$dimensional Boltzmann equation as,
\begin{equation}\label{15}
\dfrac{\partial f_q}{\partial t}+\dot{z}\dfrac{\partial f_q}{\partial z}+(\dot{p_z}+F_z)
\dfrac{\partial f_q}{\partial p_z}=C(f_q, f_g),
\end{equation}
where $F_z=-\partial_{\mu}(\delta\omega u^{\mu}u_z)$ is the mean field force derived 
from the conservation laws within EQPM as described in the Ref.~\cite{Mitra:2018akk} and 
$\dot{p_z}$ gives the electromagnetic force in the medium.
The quantity $C(f_q, f_g)$ is the collision term which quantifies the rate of change of the
momentum distribution function for different processes.

\subsection{1+1-D Boltzmann equation with the relaxation time approximation}
In order to estimate the longitudinal current density in the presence of a magnetic field, one needs to solve the 
$(1+1)-$dimensional Boltzmann equation. This can be done by employing the relaxation time 
approximation (RTA)~\cite{Mitra:2016zdw} for the collision term,
\begin{equation}\label{16}
C(f_q, f_g)=-\dfrac{\delta f_q^{(1)}}{\tau_{eff}},
\end{equation}
in which $\tau_{eff}$ is the thermal relaxation time of the process under consideration.
The Boltzmann equation within RTA takes the following form,
\begin{align}\label{17}
(v.\partial_X&+\tau^{-1}_{eff})f_{q/\bar{q}}(p_z,X)+\Bigg[\pm q_f\Big(\vec{E}(X)
+(\vec{v}\times\vec{B})\Big)\nonumber\\&+\vec{F}_{m}\Bigg].\bigtriangledown_{p_z} f_{q/\bar{q}}
 =\tau^{-1}_{eff}f^{0}_{q/\bar{q}}(p_z),
\end{align}
with $v^{\mu}=(1, \vec{v})$ and $\vec{F}_m=F_z\hat{z}$. Using the Eq.~(\ref{12}) for the longitudinal case 
(direction parallel to magnetic field) Eq.~(\ref{17}) becomes,
\begin{equation}\label{18}
(v.\partial_X+\tau^{-1}_{eff})\delta f^{(1)}_{q/\bar{q}}=(\mp q_f\vec{E}(X)-\vec{F}_m).\vec{v}{f_{q/\bar{q}}^{(0)}}^{'},
\end{equation}
in which ${f_q^{(0)}}^{'}=-\beta f_q^0(1-f_q^0)$. Since the inhomogeneity 
of the field under consideration is 
small, we have $\partial_{X}\ll \tau^{-1}_{eff}$. Hence,
\begin{equation}\label{19}
 {(v.\partial_X
+\tau^{-1}_{eff})}^{-1}\simeq \tau_{eff}(1-\tau_{eff}v.\partial_X). 
\end{equation}
We are considering the case in which inhomogeneity of the electric field is 
small so that the collision integrals are significant. 
Incorporating all the above approximations, the longitudinal current density in strong magnetic field background takes the form,
\begin{equation}\label{21}
J_{z}={J_z}_{(l)}+{J_z}_{(nl)},
\end{equation}
where ${J_z}_{(l)}$ and ${J_z}_{(nl)}$ are the linear and additional components 
due to the inhomogeneity of the electric field of the longitudinal current respectively and have the form,
\begin{widetext}
\begin{align}\label{22}
{J_z}_{(l)}&=-\sum_{l=0}^{\infty}\dfrac{q_f^2}{2\pi}\dfrac{\mid q_fB\mid}{2\pi}\mu_lN_c\int_{-\infty}^{\infty}{dp_z v_z^2
\tau_{eff}} ({f_q^{(0)}}^{'}+{f_{\bar{q}}^{(0)}}^{'})E+\delta\omega\sum_{l=0}^{\infty}\dfrac{q_f^2}{2\pi}\dfrac{\mid q_fB\mid}{2\pi}\mu_lN_c\int_{-\infty}^{\infty}{dp_z v_z^2 \dfrac{1}
{E_l} \tau_{eff}} ({f_q^{(0)}}^{'}+{f_{\bar{q}}^{(0)}}^{'})E,
\end{align}
and
\begin{align}\label{23}
{J_z}_{(nl)}&=\sum_{l=0}^{\infty}\dfrac{q_f^2}{2\pi}\dfrac{\mid q_fB\mid}{2\pi}\mu_lN_c\int_{-\infty}^{\infty}{dp_z 
v_z^2\tau^2_{eff}({f_q^{(0)}}^{'}+{f_{\bar{q}}^{(0)}}^{'})}\dot{E}-\delta\omega\sum_{l=0}^{\infty}\dfrac{q_f^2}{2\pi}\dfrac{\mid q_fB\mid}{2\pi}\mu_lN_c\int_{-\infty}^{\infty}{dp_z v_z^2\dfrac{\tau^2_{eff}}
{E_l}({f_q^{(0)}}^{'}+{f_{\bar{q}}^{(0)}}^{'})}\dot{E}\nonumber\\
&+\sum_{l=0}^{\infty}\dfrac{q_f^2}{2\pi}\dfrac{\mid q_fB\mid}{2\pi}\mu_lN_c\int_{-\infty}^{\infty}{dp_z 
\tau^2_{eff}({f_q^{(0)}}^{'}+{f_{\bar{q}}^{(0)}}^{'})}v^3_z\partial_zE-\delta\omega\sum_{l=0}^{\infty}\dfrac{q_f^2}{2\pi}\dfrac{\mid q_fB\mid}{2\pi}\mu_lN_c\int_{-\infty}^{\infty}{dp_z \dfrac{\tau^2_{eff}}
{E_l}({f_q^{(0)}}^{'}+{f_{\bar{q}}^{(0)}}^{'})}v^3_z\partial_zE.
\end{align}
\end{widetext}
In the case of momentum independent thermal relaxation time, the integral with $\partial_z\vec{E}$ vanishes 
since the integrand is an odd function of $p_z$ and hence current density will be proportional only to  
$\vec{E}$ and $\dot{\vec{E}}$ as described in the Ref.~\cite{Satow:2014lia} (only for zero chiral 
chemical potential). 
The microscopic interactions are the dynamical inputs which can be encoded through the thermal relaxation 
time $\tau_{eff}$. 
The momentum dependent thermal relaxation for the dominant 1 $\rightarrow$ 2 processes ($ k\rightarrow p+p^{'}$) 
in the magnetized medium takes the 
following form~\cite{Hattori:2016lqx,Kurian:2017yxj}, 
\begin{align}\label{23.1}
\tau_{eff}^{-1}(p_z)&=\dfrac{1}{4\omega_{l_q}}\dfrac{1}{(1-f_{q}^{0}(p_z))}\sum_{l^{'}\geq l}\int_{-\infty}^{\infty}{\dfrac{dp_{z^{'}}}{2\pi}}
\dfrac{1}{2\omega_{{l^{'}}_{\bar{q}}}}X(l, l^{'},\xi)\nonumber\\
&\times f^0_{q}(p_{z^{'}})( 1+f_{g}^{0}(p_{z^{'}}+p_z)),
\end{align}
with
\begin{equation}
\xi=\dfrac{(\omega_{l_q}+\omega_{{l^{'}}_{\bar{q}}})^2-(p_z+p_{z^{'}})^2}{2\mid q_fB\mid},
\end{equation}
and $X(l, l^{'},\xi)$ can be defined in terms of Laguerre polynomials as follows,
\begin{align}
X(l, l^{'},\xi)&=4\pi\alpha_{eff}C_2\dfrac{l!}{l^{'}!}e^{-\xi}\xi^{l^{'}-l}\Bigg[\bigg(4m^2\nonumber\\
&-4\mid q_feB\mid 
(l+l^{'}-\xi)\dfrac{1}{\xi}(l+l^{'})\bigg)F(l, l^{'}, \xi)\nonumber\\
&+16\mid q_feB\mid l^{'}(l+l^{'})\dfrac{1}{\xi}L_{l}^{(l^{'}-l)}(\xi)L_{l-1}^{(l^{'}-l)}(\xi)\Bigg].
\end{align}
where $F(l, l^{'}, \xi)=[L_{l}^{(l^{'}-l)}(\xi)]^2+\dfrac{l^{'}}{l}[L_{l-1}^{(l^{'}-l)}(\xi)]^2$ for $l>0$ 
and $F(l, l^{'}, \xi)=1$ for the lowest Landau level. Here, $\alpha_{eff}$ is the effective 
coupling within EQPM and $C_2=\dfrac{(N_c^2-1)}{2N_c}$ is the Casimir factor in which $N_c$ is the 
number of colors. The equilibrium quasigluon
distribution function is defined as~\cite{Kurian:2017yxj},
\begin{equation}\label{27}
f^0_{g}=\dfrac{z_{g}\exp{(-\beta\mid\vec{p}\mid)}}
{1+ z_{g}\exp{( -\beta\mid\vec{p}\mid} )},
\end{equation}
in which $\mid\vec{p}\mid=E_p$ for gluons. 
The longitudinal current density of magnetized QGP with HLLs can be obtained from the Eqs.~(\ref{22}) and~(\ref{23}) 
by defining the thermal relaxation time as 
in Eq.~(\ref{23.1}). 

It would be instructive to check for the LLL result in regime $T^{2}\ll \mid q_fB\mid$. 
Within the LLL approximation, we have $X(l=0, l^{'}=0, \xi)\approx 16\pi{(\alpha_{eff})}m^2C_2$ 
and the thermal relaxation time reduced to the LLL result as follows,
\begin{equation}\label{26}
\tau^{-1}_{eff}=\dfrac{2\alpha_{eff}C_2m^2}{\omega_p(1-f^{0}_q)}\dfrac{z_q}{(1+z_q)}(1+f^{0}_g(E_{p_z}))\ln (T/m).
\end{equation}
Solving Eq.~(\ref{22}) in the LLL approximation in the regime $p_{z^{'}}\sim 0$~\cite{Hattori:2016lqx,Kurian:2017yxj}, we have,
\begin{align}\label{28}
{J_z}_{(l)}&=\dfrac{q_f^2N_c}{\pi}\dfrac{\mid q_fB\mid}{2\pi}\dfrac{T}{C_2m^2\alpha_{eff}}\dfrac{h_{(l)}}{\ln(T/m)}E\nonumber\\
&-\delta\omega\dfrac{q_f^2N_c}{\pi}\dfrac{\mid q_fB\mid}{2\pi}\dfrac{1}{C_2m^2\alpha_{eff}}\dfrac{k_{(l)}}{\ln(T/m)}E,
\end{align}
where $h_{(l)}\equiv h_{(l)}(z_g,z_q)$ and $k_{(l)}\equiv k_{(l)}(z_g,z_q)$ 
describes the hot medium interactions and have the form,
\begin{align}\label{29}
h_{(l)}&=\dfrac{z_q+1}{z_q} \Big(z_g + z_q + (-z_g + z_q)\ln(z_q)\Big)\dfrac{1}{8z_q},
\end{align}
and 
\begin{align}\label{30}
k_{(l)}&=\dfrac{z_q+1}{z_q} \Big(-z_g + z_q\Big)\dfrac{1}{4z_q}.
\end{align}
The linear component ${J_z}_{(l)}$ represents the LLL Ohmic current density.  
The leading order first term of Eq.~(\ref{28}) leads to the following expression of longitudinal conductivity,
\begin{align}\label{36}
\sigma_{eff}^{L}&=\dfrac{\mid q_fB\mid}{2\pi}\dfrac{q_f^{2}N_c}{\pi C_{2}\alpha_{eff}} 
\dfrac{T}{\ln (T/m)}\dfrac{(z_{q}+1)}{2z_{q}}\nonumber\\
&\times\dfrac{1}{4}\lbrace \dfrac{(z_{q}+z_{g})-(z_{g}-z_{q})\ln(z_{q})}{z_{q} m^{2}} \rbrace,
\end{align}
which is the same form as described in the Ref.~\cite{Kurian:2017yxj}, whereas the second term describes the mean field 
contribution to LLL longitudinal current density.  As expected, the mean field contribution is more visible in 
the temperature regime not very far from the transition temperature. 

The component ${J_z}_{(nl)}$ gives 
the corrections to the longitudinal current density in terms of inhomogeneity of electric field in time and expressed as,
\begin{align}\label{31}
{J_z}_{(nl)}&=-\dfrac{q_f^2N_c}{\pi}\dfrac{\mid q_fB\mid}{2\pi}\dfrac{T^2}{C_2^2m^4\alpha^2_{eff}}
\dfrac{h_{(nl)}}{48(\ln(T/m))^2}\dot{E}\nonumber\\
&+\delta\omega_{p_z}\dfrac{q_f^2N_c}{\pi}\dfrac{\mid q_fB\mid}{2\pi}
\dfrac{T}{C_2^2m^4\alpha^2_{eff}}\dfrac{k_{(nl)}}{48(\ln(T/m))^2}\dot{E},
\end{align}
where
\begin{align}\label{32}
h_{(nl)}&=-\dfrac{(1 + z_q)}{12z^4_q}\Bigg[(z_g + z_q) \Big(-z_q (z_g + z_q) \nonumber\\
&+ 3 (z_g - z_q) (1 + z_q)\ln(1 + z_q)\Big) \nonumber\\
&+ 2 (1 + z_q) (z_g^2 - z_g z_q + z_q^2) PolyLog[2, -z_q]\Bigg],
\end{align}
and 
\begin{align}\label{33}
k_{(nl)}&=\dfrac{(1 + z_q)}{z_q}\Bigg[-3z^2_g + 3z^2_q \nonumber\\
&+2 \Big(z^2_g -z_gz_q + z^2_q\Big)\ln (z_q)\Bigg]/(24 z^3_q).
\end{align}
The second term in Eq.~(\ref{31}) defines the mean field correction to the ${J_z}_{(nl)}$. 
Note that the LLL thermal relaxation time is an even function of $p_z$  and from Eq.~(\ref{23}), 
the term with $\partial_zE$ vanishes in the LLL current density. Let us now proceed 
to discuss the case of BGK collisional kernel which could be thought of an 
improvement over RTA.  
 
\subsection{Boltzmann equation with BGK kernel in the strong magnetic field}
The effect of collisions in the magnetized hot QGP can be described 
by the Bhatnagar-Gross-Krook (BGK) collisional kernel in the 
effective (covariant) transport equation. We described Boltzmann 
equation with the collision term in the Eq.~(\ref{15}). To handle the BGK 
collisional kernel $C(f_q)$, we closely follows~\cite{krook,thoma,Li,carrington,Kumar:2017bja} and extend 
the analysis considering the extended EQPM~\cite{Kurian:2017yxj}. Here, we have,
\begin{equation}\label{40}
C(f_q)=-\nu\Bigg[f_q(p_z,X)-\dfrac{N(X)}{N^0}f^0_q(p_z)\Bigg],
\end{equation}
where, 
\begin{equation}\label{41}
f_q(p_z,X)=\bar{f}_q(p_z)+\delta f_q(p_z,X),
\end{equation}
in which $\bar{f}_q(p_z)$ is the local equilibrium part   and $\delta f_q(p_z,X)$ is the perturbed part 
of the distribution function. For the inclusion of the BGK collisional term for the equilibration of the system, while describing the 
longitudinal current density, we need to define the collisional frequency $\nu$. Here, the collisional frequency is an input 
parameter which is independent of momentum and particle species. In the current analysis, $\nu^{-1}$ 
is fixed as the thermal average of the relaxation time. Hence,
\begin{equation} 
\nu=<\tau^{-1}_{eff}>=\frac{\int{dp_z\tau_{eff}f^0_q}}{\int{d_z}f^0_q},
\end{equation} 
where $\tau_{eff}$ is the momentum dependent thermal relaxation time as defined in Eq.~(\ref{23.1}) for the 
dominant $1\rightarrow2$ process in the magnetized medium.
The particle density $N(X)$ and it's equilibrium value in the presence of the
strong magnetic field are defined as,
\begin{equation}\label{42}
N(X)=\sum_{l=0}^{\infty}\dfrac{\mid q_fB\mid}{2\pi}\mu_lN_c\int_{-\infty}^{\infty}{\dfrac{dp_z}{2\pi}f_q(p_z,X)},
\end{equation} 
and 
\begin{equation}\label{43}
N^0=\sum_{l=0}^{\infty}\dfrac{\mid q_fB\mid}{2\pi}\mu_lN_c\int_{-\infty}^{\infty}{\dfrac{dp_z}{2\pi}f^0_q(p_z)}.
\end{equation}
Note that the difference between BGK collisional term and relaxation time approximation is the rate 
$\frac{N(X)}{N^0}$. The BGK collision term is an improvement of the RTA is in the sense that it can 
conserve the particle number instantaneously. Hence, we have
\begin{equation}\label{44}
\sum_{l=0}^{\infty}\dfrac{\mid q_fB\mid}{2\pi}\mu_l\int_{-\infty}^{\infty}{\dfrac{dp_z}{2\pi}C(f_q)}=0.
\end{equation}
The $(1+1)-$dimensional relativistic transport equation of the
single quasi-particle distribution function with the BGK kernel is
given by the following equation,
\begin{align}\label{45}
&v.\partial_X\delta f_{q/\bar{q}}(p_z,X)+\Big(\pm q_f\vec{E}(X)+\vec{F}_m\Big).\bigtriangledown_{p_z} 
\bar{f}_{q/\bar{q}}=\nonumber\\
&-\nu\Bigg[\bar{f}_q(p_z)+\delta f_{q/\bar{q}}(p_z,X)-\Big(1+\dfrac{\int_{p_z}{\delta f_{q/\bar{q}}(p_z,X)}}
{N^0}\Big)f^0_q\Bigg].
\end{align}
Solving Eq~(\ref{45}) for $\delta f_{q/\bar{q}}(p_z,X)$ in the longitudinal direction yields the following form,
\begin{align}\label{46}
\delta f_{q/\bar{q}}(p_z,X)&=\Bigg[i\nu \Big( f_{q/\bar{q}}^0(p_z)-\bar{f}_{q/\bar{q}}(p_z)\Big)\nonumber\\
&+\bigg(i(\mp q_f\vec{E}(X)-\vec{F}_m).\bigtriangledown_{p_z} \bar{f}_{q/\bar{q}}(p_z)\bigg)\nonumber\\&+
i\nu f^0_q(p_z)\Bigg(\dfrac{\int_{p_z}{\delta f_{q/\bar{q}}(p_z,X)}}{N^0}\Bigg)\Bigg](iD)^{-1},
\end{align} 
in which $D=v.\partial_X+\nu$. 
Defining 
\begin{align}\label{47}
\delta {f_{q/\bar{q}}}^{(0)}&=\Bigg[\bigg(i(\mp q_f\vec{E}(X)-\vec{F}_m).\bigtriangledown_{p_z} \bar{f}_{q/\bar{q}}(p_z)\bigg)\nonumber\\
&+i\nu \Big( f_{q/\bar{q}}^0(p_z)-\bar{f}_{q/\bar{q}}(p_z)\Big)\Bigg]/(iD)^{-1},
\end{align}
we can solve the Eq.~(\ref{46}) as,
\begin{align}\label{48}
\delta f_{q/\bar{q}}(p_z,X)&=\delta {f_{q/\bar{q}}}^{(0)}+i\nu(iD)^{-1}\dfrac{f^0_q(p_z)}{N^0}\nonumber\\
&\times\Bigg[\int_{p^{'}_z}
{\delta {f_{q/\bar{q}}}^{(0)}(p^{'}_z,X)}\Bigg]
+i\nu(iD)^{-1}\dfrac{f^0_q}{N^0}\dfrac{i\nu}{N^0}\nonumber\\
&\times\Bigg[\int_{p^{'}_z}{(iD)^{-1} f^0_q(p^{'}_z)}\int_{p^{''}_z}
{\delta {f_{q/\bar{q}}}^{(0)}(p^{''}_z,X)}\Bigg]\nonumber\\
&+......  .
\end{align}
Inserting $\delta f_{q}$ and $\delta f_{\bar{q}}$ upto the first order as defined in Eq.~(\ref{48}) into the induced 
current in the Eq.~(\ref{11}), we can analytically calculate the effects of 
inhomogeneity of the electric field in the current density along with the mean field 
corrections. We have, $D^{-1}=\frac{1}{\nu}(1-\frac{v.\partial_X}{\nu})$. Hence, the leading order longitudinal 
electrical conductivity in the presence of the magnetic field $\vec{B}=B\hat{z}$ has the following form,
\begin{align}\label{49}
J_z&= -\sum_{l=0}^{\infty}q_f^2\dfrac{\mid q_fB\mid}{2\pi}\mu_lN_c\int_{-\infty}^{\infty}{\dfrac{dp_z}{2\pi}v_z^2
(D)^{-1}E ({f^0_q}^{'}+{f^0_{\bar{q}}}^{'})}\nonumber\\
& +\delta\omega\sum_{l=0}^{\infty}q_f^2\dfrac{\mid q_fB\mid}{2\pi}\mu_lN_c\int_{-\infty}^{\infty}{\dfrac{dp_z}{2\pi}\dfrac{v_z^2}{E_l}(D)^{-1}
E ({f^0_q}^{'}+{f^0_{\bar{q}}}^{'})}\nonumber\\
&-\sum_{l=0}^{\infty}\dfrac{q_f^2\nu}{N^0}\dfrac{\mid q_fB\mid}{2\pi}\mu_lN_c\int_{-\infty}^{\infty}{\dfrac{dp_z}{2\pi}v_z(D)^{-1}\Big[({f^0_q}+
{f^0_{\bar{q}}})}\Lambda\Big]\nonumber\\
&+\delta\omega\sum_{l=0}^{\infty}\dfrac{q_f^2\nu}{N^0}\dfrac{\mid q_fB\mid}{2\pi}\mu_lN_c\int_{-\infty}^{\infty}{\dfrac{dp_z}{2\pi}\dfrac{v_z}{E_l}
(D)^{-1}\Big[({f^0_q}+{f^0_{\bar{q}}})}\Delta\Big],    
\end{align}
where, 
\begin{equation}\label{50}
\Lambda=\dfrac{\mid q_fB\mid}{2\pi}\mu_l\int_{-\infty}^{\infty}{\dfrac{dp_z}{2\pi}v_z
(D)^{-1}E ({f^0_q}^{'}+{f^0_{\bar{q}}}^{'})},
\end{equation}
\begin{equation}\label{51}
\Delta=\dfrac{\mid q_fB\mid}{2\pi}\mu_l\int_{-\infty}^{\infty}{\dfrac{dp_z}{2\pi}\dfrac{v_z}{E_l}(D)^{-1}
E({f^0_q}^{'}+{f^0_{\bar{q}}}^{'})}.
\end{equation}
The second and fourth terms give the mean field contribution to the current density.
Following Eq.~(\ref{49}), the non-vanishing components of current density comes out to be in the 
form of Eq.~(\ref{21}) in which
\begin{align}\label{51.01}
&{J_z}_{(l)}= I_1, &{J_z}_{(nl)}=I_2+I_3,
\end{align}
where,
\begin{widetext}
\begin{align}\label{52}
I_1=& -\sum_{l=0}^{\infty}q_f^2\dfrac{\mid q_fB\mid}{2\pi}\mu_lN_c\int_{-\infty}^{\infty}{\dfrac{dp_z}{2\pi}
\dfrac{v^2_z}{\nu} ({f^0_q}^{'}+{f^0_{\bar{q}}}^{'})} E+\delta\omega\sum_{l=0}^{\infty}q_f^2\dfrac{\mid q_fB\mid}
{2\pi}\mu_lN_c\int_{-\infty}^{\infty}{\dfrac{dp_z}{2\pi}\dfrac{1}{\nu}\dfrac{v^2_z}{E_l} 
({f^0_q}^{'}+{f^0_{\bar{q}}}^{'})}E,\\
I_2=&\sum_{l=0}^{\infty}q_f^2\dfrac{\mid q_fB\mid}{2\pi}\mu_lN_c\int_{-\infty}^{\infty}{\dfrac{dp_z}{2\pi}
\dfrac{v^2_z}{\nu^2} ({f^0_q}^{'}+{f^0_{\bar{q}}}^{'})} \dot{E}        -\delta\omega\sum_{l=0}^{\infty}q_f^2
\dfrac{\mid q_fB\mid}{2\pi}\mu_lN_c\int_{-\infty}^{\infty}{\dfrac{dp_z}{2\pi}\dfrac{1}
{\nu^2}\dfrac{v^2_z}{E_l} ({f^0_q}^{'}+{f^0_{\bar{q}}}^{'})}\dot{E},\label{53}\\
I_3=&-\sum_{l=0}^{\infty}\dfrac{q_f^2}{N^0}\dfrac{\mid q_fB\mid}{2\pi}\mu_l\int_{-\infty}^{\infty}{\dfrac{dp_z}
{2\pi}v_z^2\Bigg[({f^0_q}+{f^0_{\bar{q}}})}
\dfrac{\mid q_fB\mid}{2\pi}\mu_l\int_{-\infty}^{\infty}{\dfrac{dp^{'}_z}{2\pi}\dfrac{1}{\nu^3} 
({f^0_q}^{'}+{f^0_{\bar{q}}}^{'})}v_z^2\partial^2 _zE\Bigg]\nonumber\\
&+\delta\omega\sum_{l=0}^{\infty}\dfrac{q_f^2}{N^0}\dfrac{\mid q_fB\mid}{2\pi}\mu_l
\int_{-\infty}^{\infty}{\dfrac{dp_z}{2\pi}\dfrac{v_z^2}{E_l}\Bigg[({f^0_q}+
{f^0_{\bar{q}}})}\dfrac{\mid q_fB\mid}{2\pi}\mu_l\int_{-\infty}^{\infty}{\dfrac{dp^{'}_z}{2\pi}
\dfrac{1}{\nu^3} ({f^0_q}^{'}+{f^0_{\bar{q}}}^{'})}v_z^2\partial^2 _zE\Bigg].
\end{align}
\end{widetext}
Since, the collision rate $\nu$ is independent of momenta, Eqs.~(\ref{52}) and~(\ref{53}) reduce to 
the momentum independent RTA result, in which the current density are 
proportional to $E$ and $\dot{E}$. The BGK kernel describes the higher order corrections to the current 
density. In the leading order we have terms proportional to $E$, $\dot{E}$ and $\partial_z^2E$. 
By defining,
\begin{equation}\label{60}
\lambda=\dfrac{i\nu}{N^0}\int_{p^{'}_z}{(iD)^{-1} f^0_q(p^{'}_z)},
\end{equation}
we can obtain the change in the distribution function for the general case 
in the following way, 
\begin{align}\label{61}
\delta f_{q/\bar{q}}(p_z,X)&=\delta {f_{q/\bar{q}}}^{(0)}\nonumber\\
&+i\nu(iD)^{-1}\dfrac{f^0_q(p_z)}{N^0}\dfrac{1}{1-\lambda}\int_{p_z}{\delta {f_{q/\bar{q}}}^{(0)}(p_z,X)},
\end{align}
in which contributions from higher order derivatives of the electric field can also be included. 
This is beyond the scope of the present work.

The background electric field in the direction of magnetic field in the RHIC has 
the following form~\cite{Satow:2014lia},
\begin{equation}\label{39}
q_f\vec{E}=\hat{z}q_fE_0z\dfrac{b}{2R}e^{\left( -\vec{X}^2/(2\sigma^2)-t/\tau_{E}\right) },
\end{equation} 
where $\sigma=4.0$ fm is the spatial width of the field, $\tau_{E}=1.0$ fm is the duration 
time of the electric field, $b=7.2$ fm is the impact factor and $R=6.38$ fm is the radius of the heavy nuclei. 
For the numerical calculations, we choose $z=4$ fm in the analysis. 
To quantify the effect of inhomogeneity of the electric field in the current density, we can define the ratio,
 \begin{equation}\label{39.01}
R_{zz}=\dfrac{{J_z}_{(l)}}{{J_z}_{(l)}+{J_z}_{(nl)}},
\end{equation} 
where ${J_z}_{(l)}$ is the leading order linear component and ${J_z}_{(l)}+{J_z}_{(nl)}$ gives the 
total longitudinal current density incorporating the additional components due to spacetime inhomogeneity 
of the field. Further, the 
longitudinal conductivity could be read off from the expression of total current density Eq.~(\ref{51.01}) by 
employing Eq.~(\ref{39}) and has the following form,
 
\begin{widetext}
\begin{align}\label{52}
\sigma_{zz}=& -\sum_{l=0}^{\infty}q_f^2\dfrac{\mid q_fB\mid}{2\pi}\mu_lN_c\int_{-\infty}^{\infty}{\dfrac{dp_z}{2\pi}
\dfrac{v^2_z}{\nu} ({f^0_q}^{'}+{f^0_{\bar{q}}}^{'})} +\delta\omega\sum_{l=0}^{\infty}q_f^2\dfrac{\mid q_fB\mid}
{2\pi}\mu_lN_c\int_{-\infty}^{\infty}{\dfrac{dp_z}{2\pi}\dfrac{1}{\nu}\dfrac{v^2_z}{E_l} 
({f^0_q}^{'}+{f^0_{\bar{q}}}^{'})}\nonumber\\
&-\sum_{l=0}^{\infty}q_f^2\dfrac{\mid q_fB\mid}{2\pi}\mu_lN_c\int_{-\infty}^{\infty}{\dfrac{dp_z}{2\pi}
\dfrac{v^2_z}{\nu^2} ({f^0_q}^{'}+{f^0_{\bar{q}}}^{'})} \dfrac{1}{\tau_E}        +\delta\omega\sum_{l=0}^{\infty}q_f^2
\dfrac{\mid q_fB\mid}{2\pi}\mu_lN_c\int_{-\infty}^{\infty}{\dfrac{dp_z}{2\pi}\dfrac{1}
{\nu^2}\dfrac{v^2_z}{E_l} ({f^0_q}^{'}+{f^0_{\bar{q}}}^{'})}\dfrac{1}{\tau_E}\nonumber\\
&+\sum_{l=0}^{\infty}\dfrac{q_f^2}{N^0}\dfrac{\mid q_fB\mid}{2\pi}\mu_l\int_{-\infty}^{\infty}{\dfrac{dp_z}
{2\pi}v_z^2\Bigg[({f^0_q}+{f^0_{\bar{q}}})}
\dfrac{\mid q_fB\mid}{2\pi}\mu_l\int_{-\infty}^{\infty}{\dfrac{dp^{'}_z}{2\pi}\dfrac{1}{\nu^3} 
({f^0_q}^{'}+{f^0_{\bar{q}}}^{'})}v_z^2(\dfrac{3}{\sigma^2}-\dfrac{z^2}{\sigma^4})\Bigg]\nonumber\\
&-\delta\omega\sum_{l=0}^{\infty}\dfrac{q_f^2}{N^0}\dfrac{\mid q_fB\mid}{2\pi}\mu_l
\int_{-\infty}^{\infty}{\dfrac{dp_z}{2\pi}\dfrac{v_z^2}{E_l}\Bigg[({f^0_q}+
{f^0_{\bar{q}}})}\dfrac{\mid q_fB\mid}{2\pi}\mu_l\int_{-\infty}^{\infty}{\dfrac{dp^{'}_z}{2\pi}
\dfrac{1}{\nu^3} ({f^0_q}^{'}+{f^0_{\bar{q}}}^{'})}v_z^2(\dfrac{3}{\sigma^2}-\dfrac{z^2}{\sigma^4})\Bigg].
\end{align}
\end{widetext}
Note that the expression of conductivity as defined in Eq.~(\ref{52}) could be possible 
due to the form of the electric field as defined 
in Eq.~(\ref{39}).  The leading order contribution to the longitudinal conductivity
 can be described from the momentum independent RTA (first four terms in Eq.~(\ref{52})). 
In contrast, the estimations with BGK kernel described the conductivity with higher order corrections 
due to the spacetime inhomogeneities of the electric field. 
We shall now proceed to investigate the temperature dependence of the ratio $R_{zz}$ in Eq.~(\ref{39.01}) and 
the longitudinal conductivity, $\sigma_{zz}$ in Eq.~(\ref{52}).
\begin{figure}[h]
  \subfloat{\includegraphics[height=6.2cm,width=9.5 cm]{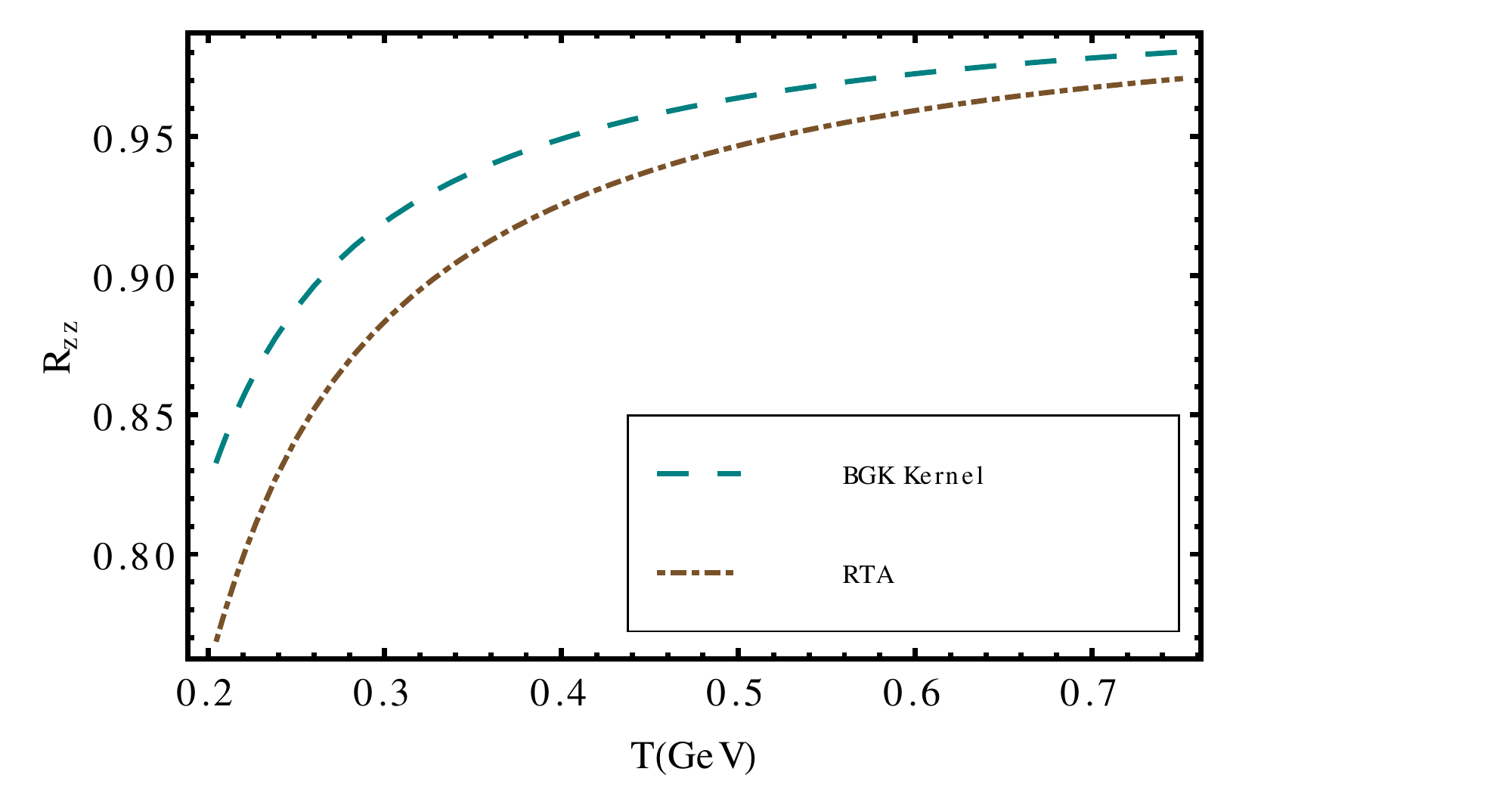}}
\caption{Temperature behaviour of the ratio $R_{zz}$ at 
$\mid q_fB\mid=10 m^2_{\pi}$ for RTA and BGK collision kernels considering upto 20 LLs.}
\label{f1}
\end{figure} 

\begin{figure}
\hspace{-.6cm}
 \subfloat{\includegraphics[height=5.9cm,width=7.5cm]{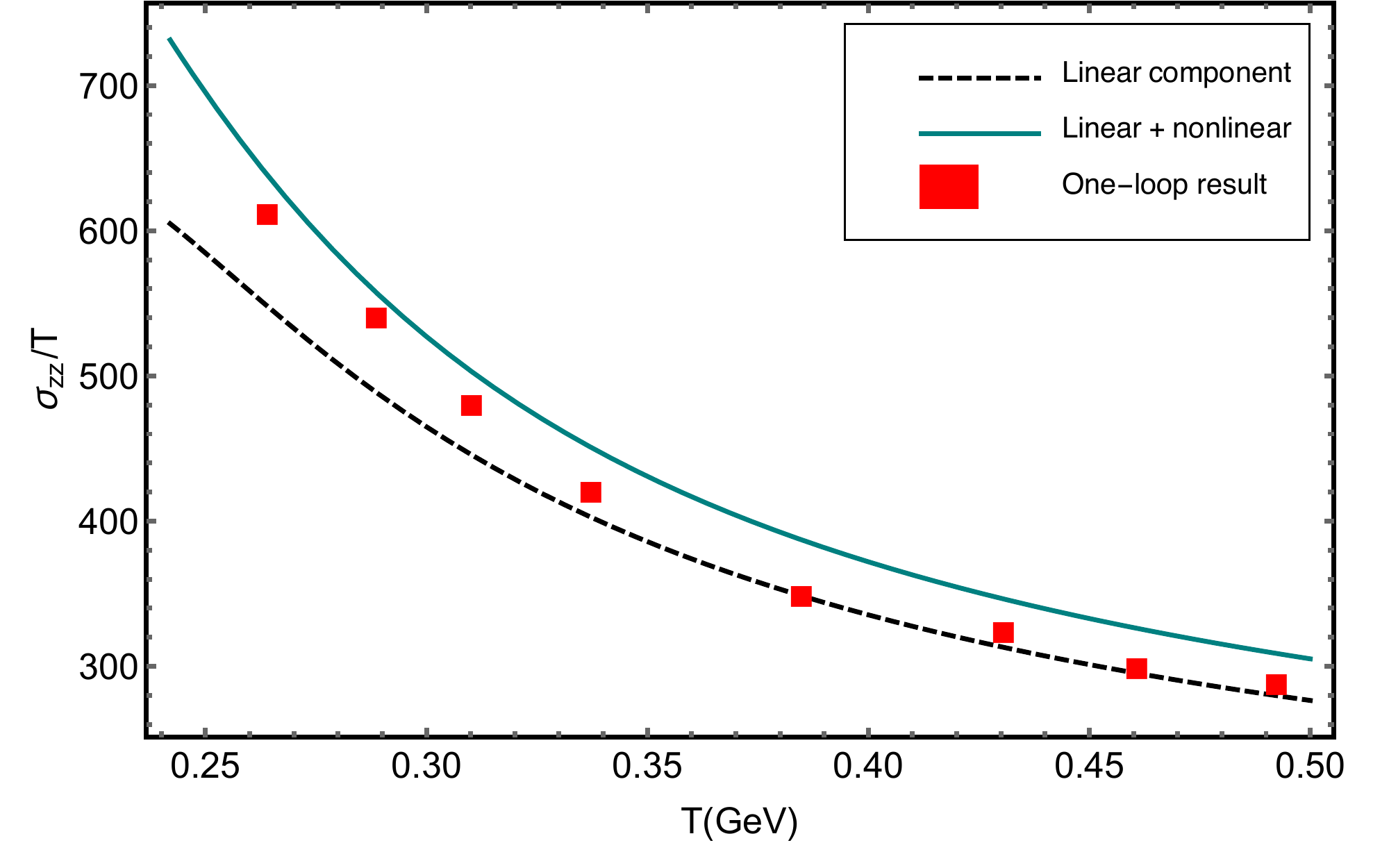}}
\caption{Behaviour of $\sigma_{zz}/T$ in the LLL approximation of 
quarks/antiquarks as a function of temperature 
at $\mid q_fB\mid=10 m^2_{\pi}$ within the RTA.  One-loop LLL
result for the linear component has been taken from~\cite{Hattori:2016cnt} for 
the same value of the magnetic field.}
\label{f2}
\end{figure}

\section{Results and discussions}
We initiate our discussion with the temperature behaviour of the longitudinal current density of 
the hot magnetized QGP in the inhomogeneous electric field.
We have estimated the longitudinal current density for both 
RTA and BGK collision kernels in the presence of the magnetic field considering the HLL contributions. 
The temperature behaviour of the ratio of 
the linear component of longitudinal current density 
to the total longitudinal current density, $R_{zz}$
as described in the Eq.~(\ref{39.01}) is depicted in the 
Fig.~\ref{f1}. The ratio $R_{zz}$ 
is plotted for both RTA and BGK collision terms and we observe a similar behaviour for both the 
collision integrals.  The $R_{zz}$ increase with increasing temperature and seen to saturate at 
higher temperatures. The temperature 
behaviour of the ratio $R_{zz}$ in the presence of the magnetic field indicates 
that the effects of spacetime inhomogeneity of electric field are significant in the temperature regime 
near to the transition temperature $T_c$ 
whereas the effects are negligible at high temperatures.
\begin{figure*}
 \centering
 \subfloat{\includegraphics[height=6.50cm,width=9.42cm]{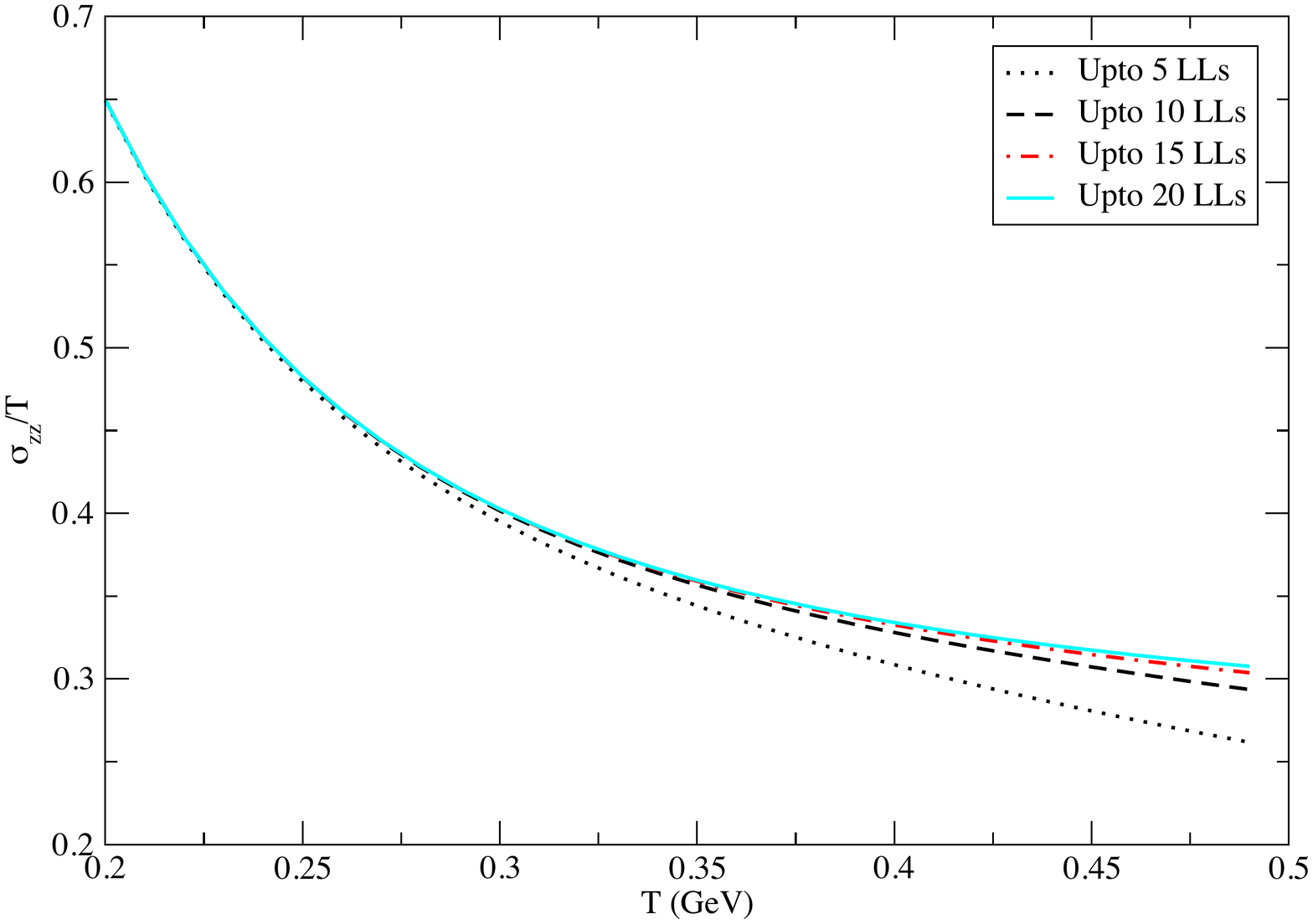}}
 \hspace{-12mm}
 \subfloat{\includegraphics[height=6.50cm,width=9.55cm]{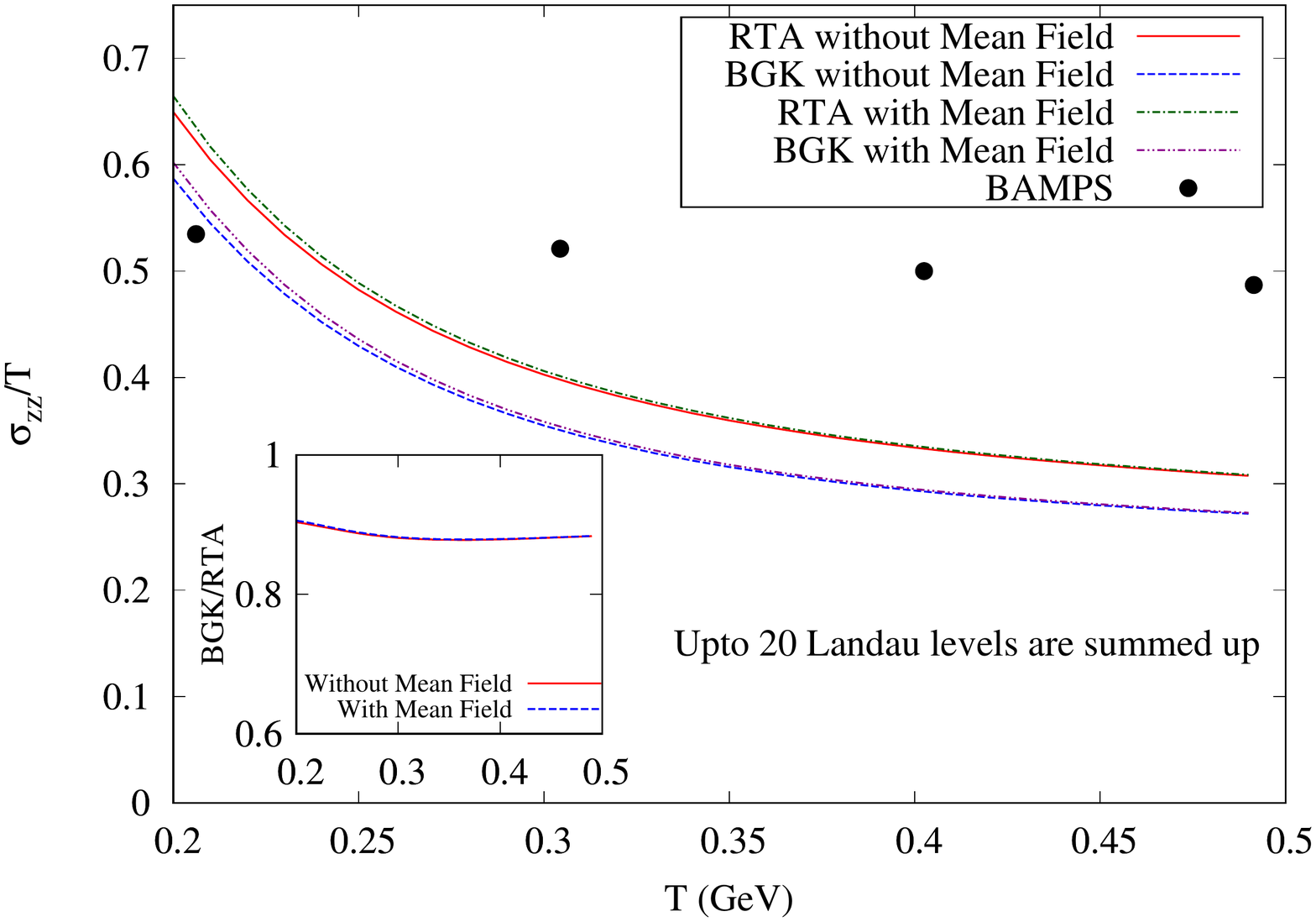}}
\caption{(Left panel) The effect of HLLs on the temperature behaviour of $\sigma_{zz}/T$ 
at $\mid q_fB\mid=10 m^2_{\pi}$. (Right panel) Mean field corrections to the $\sigma_{zz}/T$ 
with the RTA and BGK collision integrals considering up to 20 LLs. 
The results are also compared with the BAMPS estimation at $B=0$~\cite{Greif:2014oia}. }
\label{f3}
\end{figure*}

The hot medium effects are embedded in the quark and gluonic effective fugacity quasiparton 
distribution functions as well as in the modified part of dispersion relation. The mean field term of 
the effective covariant kinetic theory employed in the current analysis involves the fugacity parameters and 
it's derivatives. The mean field force terms $F^i=\partial_{\mu}(\delta\omega u^{\mu}u^{i})$ are of the second 
order in gradient since $\delta\omega$ itself is a temperature gradient of the effective fugacity ($z_{g/q}$). 
Note that at high 
temperature regime, the effective fugacity varies very slowly with temperature and hence the mean field 
effects are negligible in that regime. 

In the LLL approximation, we have $T^{2}\ll \mid q_fB\mid$.  In this case, the  linear component 
of the current density in the RTA reduced to the form as defined in Eq.~(\ref{28}).
The first term in the Eq.~(\ref{28}) exactly gives back the leading order contribution of longitudinal current 
density for $1\rightarrow 2$ processes in the strong magnetic field as estimated in the 
Refs.~\cite{Kurian:2017yxj,Hattori:2016lqx}, whereas 
the second term describes the mean field contribution to the linear component of the current density. In the 
similar way, the Eq.~(\ref{31}) describes the additional component 
of the current density due to the spacetime inhomogeneity of the field 
of the magnetized QGP. We compared these results with 
another approach for the calculation of leading order longitudinal conductivity from one-loop quantum field theory in Fig.\ref{f2}. 
We observe that $\sigma_{zz}/T$ decrease with increasing temperature.
The authors of the Ref.~\cite{Hattori:2016cnt} provide the one-loop result of the linear component 
of longitudinal conductivity in the strong magnetic field arising 
from $1\rightarrow 2$ processes of which kinematics are satisfied by fermion dispersion relation with ideal 
EoS. The quantitative difference in the temperature behaviour 
of the linear component of longitudinal conductivity around the regime 
closer to the transition temperature $T_c$ reveals the effect of hot QCD 
medium interactions. The effects of the inhomogeneity of the electric field to the longitudinal 
conductivity of magnetized QGP 
are also visible near $T_c$. The LLL results in  Eq.~(\ref{28}) and Eq.~(\ref{31}) have an enhancement on the limit $m\rightarrow 0$ 
as described in~\cite{Hattori:2016lqx, Hattori:2016cnt}. 

 In the weakly coupled regime 
$gT\ll \sqrt{\mid q_fB\mid}$, the HLL contributions have significant effect in the 
transport coefficients~\cite{Kurian:2018qwb, Fukushima:2017lvb}. 
The temperature behaviour of $\sigma_{zz}/T$  with HLL contributions are 
depicted in the Fig.~\ref{f3} (left panel). Quantitatively, the  $\sigma_{zz}/T$ with the HLL effects remains in the range 
of lattice data results $0.1\le\sigma/T\le 1.0$. This observation is quantitatively 
consistent with the result of the longitudinal conductivity with HLL 
contribution, $\sigma_{zz}/T\approx 0.7$ at $T=200$ MeV where 
$\mid q_fB\mid=10 m^2_{\pi}$ and baryon chemical potential $\mu=0$, 
as described in the recent work~\cite{Fukushima:2017lvb}. 
The mean field effect induces visible modifications to 
the longitudinal current density of the hot magnetized QCD matter at the low temperature regimes as 
plotted in Fig.~\ref{f3} (right panel). 
It is to be observed that for the temperatures above $300$MeV, the mean field induced corrections are extremely mild. 
Their inclusion is essential to 
describe the behaviour of the current density in the temperatures closer to the transition temperature $T_c$. We observe 
similar temperature behaviour of $\sigma_{zz}/T$ for both RTA and BGK collision kernel. The quantitative 
difference in the results with RTA and BGK are depicted in Fig.~\ref{f3} (right panel). We have also compared the
HLL result for $\sigma_{zz}/T$ with the BAMPS estimation at $B=0$~\cite{Greif:2014oia}. 

Finally, we have observed that the effects of inhomogeneity of the electric field on $\sigma_{zz}$ 
are significant in the lower temperature 
regimes. The EoS dependence of the longitudinal conductivity is more visible near to the transition 
temperature. Also, note that the HLL contributions to the conductivity are significant 
in the weakly coupled regime.

\section{Conclusion and Outlook}
In conclusion, we have obtained the temperature behaviour of the 
longitudinal current density and the conductivity of the hot interacting QCD 
matter in the presence of the magnetic field. We considered 
the effects of spacetime inhomogeneity of the electric field and the contribution of higher Landau levels in the analysis. 
We have incorporated the hot QCD medium interactions by exploiting the quasiparticle description of 
quark and gluonic degrees of freedom.
We have estimated the longitudinal conductivity for both RTA and BGK collision term in the magnetized QGP medium.
Setting up a $(1+1)-$dimensional effective covariant kinetic theory with proper collision integral defines the 
mean field force term which indeed appears as the mean field corrections to the longitudinal current density and conductivity.
The mean field contributions appeared to be significant in the vicinity of the transition temperature $T_c$. The above 
observation is in line with the earlier results for transport coefficients~\cite{Kurian:2018qwb}. We employed RTA for the 
computation of the longitudinal current density for the  $1\rightarrow 2$ 
processes which are dominant in the presence of the magnetic field. Furthermore, 
the current density and conductivity with the BGK collision kernel have been computed in the magnetic field 
followed by the comparison of the results with that of RTA. 

The effects of 
inhomogeneity of the electric field to the longitudinal 
current density and the conductivity are seen to be quite significant at 
the temperature regime near to the transition temperature $T_c$. The 
ratio $R_{zz}$ varies between $75\% - 95\%$ in the temperature range $200$ MeV$-750$ MeV for both RTA and BGK 
collision kernels. 
Notably, in the weakly coupled regime $gT\ll \sqrt{\mid q_fB\mid}$, the inclusion 
of higher Landau level contribution is essential in the 
estimation of longitudinal conductivity of the magnetized medium.
Finally, both the mean field corrections and the effects of electric field inhomogeneity seen to have 
significant impact on the longitudinal current density and the electrical 
conductivity in the presence of the strong magnetic field.

We intend to estimate the electromagnetic responses of the chiral plasma with the mean field 
contribution in the near future. In addition, developing the second order dissipative relativistic
hydrodynamics for the magnetized QGP medium from transport theory with the EQPM would 
be another interesting direction to work.

\section*{acknowledgments}
V.C. would like to acknowledge Science 
and Engineering Research
Board (SERB), Govt. of India for the Early
Career Research Award (ECRA/2016) and 
INSA-DST for the INSPIRE Faculty Fellowship (IFA-13/PH-55). 
We would like to thank Hemanth H. and Snigdha Ghosh for 
providing numerical help in the project. 
We also record our gratitude to the people of India 
for their generous support 
for the research in basic sciences.

\appendix
\section{LLL Longitudinal conductivity within RTA at $\mu\neq 0$ }\label{A}
The extension 
of the EQPM to finite baryon/quark chemical potential 
$\mu$ in the presence of the strong magnetic field is 
quite straightforward and has the form,
\begin{equation}\label{3}
f^0_{q/\bar{q}}=\dfrac{z_{q}\exp{[-\beta (\sqrt{p_{z}^{2}+m^{2}
}\mp\mu)]}}{1+ z_{q}\exp{[-\beta (\sqrt{p_{z}^{2}+m^{2}
}\mp\mu)]}}.
\end{equation}
Here, $z_g$, $z_q$ are not related with any conserved 
number current in the hot QCD medium. They have been 
merely introduced to encode the hot QCD medium
effects in the EQPM.
Following same prescriptions in the derivation of relaxation time as in the Ref.~\cite{Kurian:2018dbn}, $\tau_{eff}$ for 
the $1\rightarrow 2$ processes with finite chemical potential $\mu$ can be expressed as,
\begin{widetext}
\begin{equation}
\tau^{-1}_{eff}=\dfrac{2\alpha_{eff}}{\omega_p(1-f^{(0)}_q)}\dfrac{z_q}{(\exp{(\dfrac{\mu}
{T})}+z_q)}(1+f^{(0)}_g(E_p))\ln (T/m),
\end{equation}
in which $\alpha_{eff}(T,\mu,\mid q_fB\mid)$ is the effective coupling with finite chemical 
potential and has the following form~\cite{Kurian:2018qwb},
\begin{align}
\dfrac{\alpha_{eff}}{\alpha_{s}(T)}=\dfrac{\dfrac{6T^{2}}
{\pi^{2}}PolyLog[2,z_{g}]
+\dfrac{3\mid q_fB\mid}{\pi^{2}}\Bigg(\dfrac{z_{q}}
{(1+z_{q})}+\dfrac{\mu^2}{2T^2}\dfrac{(z_q-z^2_q)}
{(1+z_q)^3}\Bigg)}{ T^{2}+\dfrac{3\mid q_fB\mid}{2\pi^{2}}}.
\end{align}
Solving Eq.~(\ref{22}) with finite chemical potential, we have,
\begin{align}
{J_z}_{(l)}&=-\dfrac{q_f^2}{2\pi}\dfrac{\mid q_fB\mid}{(2\pi)}\dfrac{T}{\ln(T/m)}h_{(l)}E,
&\text{and}
&&{J_z}_{(nl)}=\dfrac{q_f^2}{2\pi}\dfrac{\mid q_fB\mid}{(2\pi)}\dfrac{T^2}{48(\ln(T/m))^2}h_{(nl)}(\dot{E}+\partial_zE),\\
\end{align}
\vspace{.2cm}
where the $h_{(l)}(z_g,z_q,\mu/T)$ and $h_{(nl)}(z_g,z_q,\mu/T)$  describe the hot medium interactions and have the form,
\begin{align}
h_{(l)}&=\dfrac{(1+z_qe^{-\mu/T})}{\alpha_{eff}}\Bigg\{ \Big[(-z_g-z_q e^{\mu/T})\Big]+
\Big[(-z_g+z_q e^{\mu/T})\bigg(\ln(1+\dfrac{e^{-\mu/T}}{z_q})
-\ln(1+z_qe^{\mu/T})\bigg)\Big]\nonumber\\
&-\Big[e^{2\mu/T}
\bigg(z_g+z_q e^{-\mu/T}
-(-z_g+z_q e^{-\mu/T})
\Big(\ln(1+\dfrac{e^{\mu/T}}{z_q})
-\ln(1+z_q e^{-\mu/T})\Big)\bigg)\Big]
\Bigg\} 
\dfrac{1}{8z_q^2m^2},
\end{align}
and
\begin{align}
h_{(nl)}&=\Bigg[\Bigg\{e^{-4\mu/T}(z_q+ e^{\mu/T})^2
\Big[(z_q+z_g e^{\mu/T})
\bigg(z_q+z_g e^{\mu/T}+3(z_g e^{\mu/T}
-z_q)\ln(1+\dfrac{e^{\mu/T}}{z_q} )
+(3z_q-3z_g e^{\mu/T})\nonumber\\
&\times\ln(1+z_q e^{-\mu/T})\bigg)-2(z^{2}_{q}
+z^{2}_{g} e^{2\mu/T}-z_q z_g e^{\mu/T})\Big(PolyLog[2,-\dfrac{ e^{\mu/T}}{z_q}]
+PolyLog[2,- z_q e^{-\mu/T}]\Big)\Big]\Bigg\}\nonumber\\
&+\Bigg\{ (1+ z_q e^{\mu/T})^2
\Big[(z_g+z_q e^{\mu/T})\bigg(z_g+z_q e^{\mu/T}
+3(z_g -z_q e^{\mu/T})\ln(1+ z_q^{-1}e^{-\mu/T})\bigg)
+3(z_g-z_q e^{\mu/T})\nonumber\\
&\times\ln(1+z_q e^{\mu/T})
-2(z^{2}_{q} e^{2\mu/T}
+z^{2}_{g} -z_q z_g e^{\mu/T})\Big(PolyLog[2,-\dfrac{ e^{-\mu/T}}{z_q}]
+PolyLog[2,- z_q e^{\mu/T}]\Big)\Big]  \Bigg\}\Bigg]\dfrac{(z_q+ e^{\mu/T})^2}{\alpha^2_{eff} z_q^6 m^4}.
\end{align}
\end{widetext}

{}

\end{document}